\documentclass[11pt]{amsart}
\reversemarginpar
\newcommand{\commentout}[1]{}

\usepackage{amsmath}
\usepackage{amssymb}

\newcommand{\nwc}{\newcommand}

 \relax
\newif\ifatten\attenfalse
 \relax
\newcommand{\atten}[1]{%
  \unskip
    \ifatten
    \raisebox{2pt}{\ \colorbox{yellow}{\hspace{.25cm}}}
      \fi
      }

\newcommand{\lt}{\left}
\newcommand{\rt}{\right}
\newcommand{\vas}{\varepsilon}
\newcommand{\lan}{\left\langle}
\newcommand{\ran}{\right\rangle}
\newcommand{\tvas}{T_t^\vas}

\newcommand{\veptil}{\tilde{V}_t^\vas}
\newcommand{\vep}{{V}_t^\vas}
\newcommand{\cv}{{\ml V}^\ep_t}
\newcommand{\cvtil}{\tilde{{\ml V}}^\ep_t}
\newcommand{\n}{\nabla}
\newcommand{\tkappa}{\tilde\kappa}

\nwc{\nwt}{\newtheorem}

\nwt{prop}{Proposition}
\nwt{proposition}{Proposition}
\nwt{lemma}{Lemma}
\nwt{theorem}{Theorem}
\nwt{cor}{Corollary}
\nwt{remark}{Remark}
\nwt{definition}{Definition} 

\nwc{\bal}{\begin{align}}
\nwc{\be}{\begin{equation}}
\nwc{\ben}{\begin{equation*}}
\nwc{\bea}{\begin{eqnarray}}
\nwc{\beq}{\begin{eqnarray}}
\nwc{\bean}{\begin{eqnarray*}}
\nwc{\beqn}{\begin{eqnarray*}}
\nwc{\beqast}{\begin{eqnarray*}}

\nwc{\eal}{\end{align}}
\nwc{\ee}{\end{equation}}
\nwc{\een}{\end{equation*}}
\nwc{\eea}{\end{eqnarray}}
\nwc{\eeq}{\end{eqnarray}}
\nwc{\eean}{\end{eqnarray*}}
\nwc{\eeqn}{\end{eqnarray*}}
\nwc{\eeqast}{\end{eqnarray*}}

\nwc{\ep}{\varepsilon}
\nwc{\vrho}{\varrho}
\nwc{\orho}{\bar\varrho}
\nwc{\ou}{\bar u}
\nwc{\vpsi}{\varpsi}
\nwc{\lamb}{\lambda_\varepsilon}

\nwc{\nn}{\nonumber}
\nwc{\bm}{\boldmath}
\nwc{\mf}{\mathbf}
\nwc{\ml}{\mathcal}

\nwc{\IA}{\mathbb{A}} 
\nwc{\IB}{\mathbb{B}}
\nwc{\IC}{\mathbb{C}} 
\nwc{\ID}{\mathbb{D}} 
\nwc{\IM}{\mathbb{M}} 
\nwc{\IP}{\mathbb{P}} 
\nwc{\II}{\mathbb{I}} 
\nwc{\IE}{\mathbb{E}} 
\nwc{\IF}{\mathbb{F}} 
\nwc{\IG}{\mathbb{G}} 
\nwc{\IN}{\mathbb{N}} 
\nwc{\IQ}{\mathbb{Q}} 
\nwc{\IR}{\mathbb{R}} 
\nwc{\IT}{\mathbb{T}} 
\nwc{\IZ}{\mathbb{Z}} 

\nwc{\cE}{{\ml E}}
\nwc{\cP}{{\ml P}}
\nwc{\cL}{{\ml L}}
\nwc{\cR}{{\ml R}}
\nwc{\cV}{{\ml V}}
\nwc{\cC}{{\ml C}}
\nwc{\cA}{{\ml A}}
\nwc{\cK}{{\ml K}}
\nwc{\cB}{{\ml B}}
\nwc{\cD}{{\ml D}}
\nwc{\cF}{{\ml F}}
\nwc{\cM}{{\ml M}}
\nwc{\cG}{{\ml G}}
\nwc{\cH}{{\ml H}}

\begin{document}

\title{Invariance Principle for Inertial-Scale Behavior of Scalar Fields
in Kolmogorov-type Turbulence}

\author{Albert C. Fannjiang}

\thanks{Department of Mathematics,
University of California at Davis,
Davis, CA 95616
Internet: fannjian@math.ucdavis.edu
}

\maketitle

\bigskip
\centerline{\small Dedicated to George Papanicolaou on the occasion
of his 60th birthday}
\bigskip

\begin{abstract}
We prove limit theorems for small-scale pair dispersion in
synthetic velocity fields with  power-law spatial spectra  and wave-number
dependent correlation times. These limit theorems
are related to 
a family of generalized Richardson's laws with a limiting case
corresponding to Richardson's $t^3$ and $4/3$-laws.
We also characterize a regime 
of positive dissipation of passive scalars.
\end{abstract}

\section{Introduction}

The celebrated Richardson's  $t^3$-law \cite{Ri} states that a pair of particles
located at $(x^{(0)}(t),x^{(1)}(t))\in \IR^{2d}$
being transported in
the incompressible turbulence 
satisfies
\be
\label{rich}
\IE |x^{(1)}(t)-x^{(0)}(t)|^2 \approx C_R \bar{\ep}t^3\quad\hbox{for}\,
\,
\ell_1\ll |x^{(1)}(t)-x^{(0)}(t)|\ll \ell_0
\ee
where $\bar{\ep}$ is the energy dissipation rate, $C_R$ is
the Richardson constant and $\ell_0$ and $\ell_1$ are
respectively the integral and viscous scales.
Here and below $\IE$ stands for the expectations
w.r.t. the ensemble of the velocity fields.
This law has been confirmed experimentally (\cite{JPT}, \cite{Ta}, \cite{MY})
and numerically (\cite{ZB}, \cite{EM}, \cite{FV}, \cite{BCCV}).
A stronger statement is
that the relative diffusivity of the tracer particles
is proportional
to the $4/3$ power of their momentary separation, and this
is called Richardson's $4/3$-law (\cite{Ri}. See also
\cite{Ob}, \cite{Co}, \cite{Ba}, \cite{Lin}).
This paper presents  several small-scale limit
theorems (Theorem \ref{thm1}, \ref{thm2} and \ref{thm3}) 
related to the Richardson's laws
for
a family of colored-noise-in-time
velocity fields that have Kolmogorov-type spatial spectra
and wave-number dependent correlation times.
The other aspect of the scaling limit concerns the
dissipation of the scalar field in the limit of
vanishing molecular diffusion (Corollary 1 and 2).

The nature of time correlation
in fully developed turbulences in the inertial range is not
entirely clear (see \cite{MK}
and the references therein). But it seems reasonable to assume
that, to the leading order, 
the temporal correlation structure of
the {\em Eulerian} velocity field $u(t,x)$ is determined by the
energy-
containing velocity components
above the integral scale, consistent
with Taylor's hypothesis commonly used in the
fluid flow measurements in the presence of a mean flow
or the random sweeping hypothesis in
the absence of a mean flow (see 
\cite{Ten} and \cite{Pr}). In both cases the
temporal correlation function on the small scales
is anisotropic and depends on external forcing.
The more robust features of small scale turbulence 
can 
be revealed by considering
the relative
velocity field $U(t,x)=u(t,x+x^{(0)}(t))-u(t,x^{(0)}(t))$, with respect to
a reference fluid particle
$x^{(0)}(t)$, which tends to preserve invariance properties
of the fluid equations.
The velocity field $u(t,x+x^{(0)}(t))$ as viewed
from a fluid particle, which is 
a useful tool
for turbulence modeling
\cite{BL}, \cite{LPP},
is called the quasi-Lagrangian velocity field
in the physics literature 
and is an example of
the general notion of the Lagrangian environment
process \cite{Os}, \cite{PV},
\cite{FK}.

We assume \cite{phd}, \cite{FKP}  that
the two-time structure function of $U(t,x)$ 
has the  
power-law form
\beq
\label{ou}
&&\IE [U(t,x)-U(t,y)]\otimes[U(s,x)-U(s,y)]\\
&&=\int_{\IR^d}2[1-\cos{(k\cdot (x-y))}]
\exp{(-a|k|^{2\beta}|t-s|)}
\cE_{(\ell_1,\ell_0)}
(\alpha, k) |k|^{1-d}d k,\quad\alpha\in(1,2),\quad\beta>0,\,\,a>0
\nonumber
\eeq
with the energy spectrum
\be
\label{power}
\cE_{(\ell_1,\ell_0)}(\alpha, k)=\left\{\begin{array}{ll}
E_0(\II-k\otimes k|k|^{-2})|k|^{1-2\alpha},&\hbox{for}\,\,|k|\in (\ell_0^{-1},
\ell_1^{-1})\\
0,& \hbox{for}\,\, |k|\not\in (\ell_0^{-1},\ell_1^{-1})
\end{array}
\rt.
\quad \ell_0<\infty,\,\, \ell_1>0,\,\,E_0>0
\ee
where $\ell_1, \ell_0$ are respectively the viscous and integral scales.
The assumed temporally stationary vector field $U(t,x)$
has homogeneous spatial increments and its
 expectation $\IE_s[U(t,x)]$,
  conditioning on the events up to time $s<t$, is assumed to
  admit the spectral representation
  \beq
  \nonumber
  \lefteqn{\IE_s[U(t,x)-U(t,y)]}\\
  & = &\int_{\IR}[1-\exp{(i k\cdot (x-y))}]
  \exp{(-a|k|^{2\beta}|t-s|)}
  \hat{U}(s, dk),\quad s<t
  \label{ou2}
  \eeq
  where $\hat{U}(t,k)$ is a time-stationary process
  with uncorrelated increments over $k$ such that
  \be
  \label{8}
  \IE[\hat{U}(t,dk)\hat{U}^*(t,dk')]=
  \cE_{(\ell_1,\ell_0)}(\alpha,k)\delta(k-k')\,\, dkdk',\quad\forall t, k, k'.
  \ee
 The exponential form of the temporal correlation in (\ref{ou}) and
 (\ref{ou2}) is not important for us; it can be replaced by
 a more general one like 
 \[
 \rho (a|k|^{2\beta}|t-s|)
 \]
 with an integrable function $\rho(\tau)$ decaying to zero as $\tau\to \infty$.
 Since the exponential form seems to agree well with 
 the Lagrangian measurements (see \cite{SY} for the Reynolds number
 around $100$ and \cite{VD} for high Reynolds numbers) we will use it
 for the sake of simplicity.

Set the rescaled velocity
\be
\label{rescale}
U_\lambda(t,x)\equiv 
\lambda^{1-\alpha}U
(\lambda^{2\beta}t,\lambda x). 
\ee
 Then $U_\lambda(t,x)$
 \commentout{
 is a temporally stationary velocity
 field with homogeneous spatial increments
 such that
 \beq
 \nonumber
 \lefteqn{\IE_s[V(t,x)-V(t,y)]}\\
 & = &\int_{L^{-1}< |k|<K}[1-\exp{(i k\cdot (x-y))}]
 \exp{(-a|k|^{2\beta}|t-s|)}
 \hat{V}(s, dk),\quad s<t
 \label{ou22}
 \eeq
 where $\hat{V}(t,k)$ is a time-stationary process
 with correlated increments over $k$ such that
 \be
 \label{8'}
 \IE[\hat{V}(t,dk)\hat{V}^*(t,dk')]=
 \cE_{(\ell_1\lambda^{-1},\ell_0\lambda^{-1})}
 (\alpha,k)\delta(k-k')\,\, dkdk',\quad\forall t, k, k'
 \ee
 where
 }
 has the energy spectrum
 \be
 \label{8''}
 \cE_{(\ell_1\lambda^{-1},\ell_0\lambda^{-1})}(\alpha,k)=
 \left\{\begin{array}{ll}
 E_0(\II-k\otimes k|k|^{-2})
 |k|^{1-2\alpha},&\hbox{for}\,\,|k|\in (\ell_0^{-1}\lambda,
 \ell^{-1}_1\lambda)\\
 0,& \,\, \hbox{else}
 \end{array}
 \rt.
 \ee
However, we do {\em not} assume in this paper the full scale-invariance,
namely, 
\be
U_\lambda(t,x) \stackrel{\rm d}{=}
U(t,x),\quad\hbox{for}\,\,\ell_1=0,\,\,\ell_0=\infty
\label{self}
\ee
where $\stackrel{\rm d}{=}$ means the identity of the distributions.
Instead, we assume the weaker assumption
of the 4th order scale invariance, i.e.
that up to the 4th moments of the velocity field can be estimated
in term of the energy spectrum as in the case of
Gaussian fields.

The viscous and integral scales $\ell_1,\ell_0$
can be related to each other via the 
Reynolds number $\mbox{Re}$
as
\[
\frac{\ell_0}{\ell_1}\sim \mbox{Re}^{\frac{1}{4-2\alpha}}
\]
by using
the positivity of kinetic energy dissipation of fluid in
the limit $\mbox{Re}\to \infty$.
The  correlation time $a^{-1}|k|^{-2\beta}$
decreases as the wave number $k$ increases.
The spatial Hurst exponent of the velocity equals $\alpha-1$
in the inertial range $(\ell_1,\ell_0)$.
It should be noted that 
because of the temporal stationarity of
the Lagrangian field $u(t,x+x^{(0)}(t))$ \cite{FK}, \cite{Z}
$U(t,x)$ has the same one-time statistics 
as the Eulerian velocity $u(t,x)$; in particular they share
the same energy spectrum, but their multiple-time
statistics are usually different.
We could work with the modified von Karman spectrum but it is
irrelevant for our purpose since we are concerned with
transport in the inertial-convective range.

\commentout{The parameter $\beta$ can then be determined from
the so called Lagrangian velocity correlation
\be
\label{lag}
\IE[u(t,x^{(0)}(t))u(0,x^{(0)}(0)]\approx C_L
(\frac{v_0}{\ell_0})^{(\alpha-1)/\beta}t^{(\alpha-1)/\beta}
\ee
in the temporal inertial range where  
\[
C_L=E_0\ell_0^{2\alpha-2}c_0^{(\alpha-1)/\beta}
\]
is the Lagrangian Kolmogorov constant.
Application of  Kolmogorov's similarity hypothesis to
Lagrangian velocity $u(t,x^{(0)}(t))$ in the inertial range
leads to $C_L\bar{\epsilon}t$ on the right side  of (\ref{lag})
(see \cite{MY}).
Consequently $\alpha-\beta=1$
and $\beta=1/3$ for $\alpha=4/3$.
}

It is convenient to express the coefficients $E_0, a$ 
in terms of $U_0$, the root mean square {\em longitudinal}
velocity increment over the integral length $\ell_0$,
as
\be
\label{const22}
E_0\approx C_\alpha U_0^2\ell_0^{2-2\alpha},\quad
a\approx c_0\ell_0^{2\beta-1}U_0,\quad \hbox{as}\,\,\ell_0/\ell_1\to\infty
\ee
with dimensionless constants $c_0$ and
\beq
\label{cd}
C_\alpha=\frac{(4\pi)^{d/2}2^{2\alpha-3}(2\alpha-2)\Gamma(\alpha+d/2)}{(d-1)\Gamma(2-\alpha)}
\eeq
where $\Gamma(r)$ is the Gamma function.

Assuming that the lifetime (i.e. correlation time $\tau(k)=a^{-1}|k|^{-2\beta})$
of eddy of size $|k|^{-1}$  is same as its turnover time one gets the relation
\be
\label{r.1}
\alpha+2\beta=2.
\ee
Assuming that the energy flux given by
$\cE_{(\ell_1,\ell_0)} |k|/\tau(k)$ is constant across the scales in
the inertial range one gets the relation
\be
\label{r.2}
\alpha-\beta=1.
\ee
The values of parameters satisfying
both eq. (\ref{r.1}) and (\ref{r.2}) correspond to the 
Kolmogorov spectrum with $\alpha=4/3,\beta=1/3$.
For the Kolmogorov spectrum, one has
the expression, by estimating $\bar{\ep}$ by $U_0^3\ell_0^{-1}$,
\be
\label{const}
E_0\approx C_\alpha
\bar{\ep}^{2/3},\quad a\approx c_0\bar{\ep}^{1/3}.
\ee

Writing $x(t)=x^{(1)}(t)-x^{(0)}(t)$  and adding the molecular diffusivity
$\kappa$
we have the following It\^o's stochastic  equation
for the pair separation $x(t)$
\beqn
d x(t)&=&[u(t,x^{(0)}(t)+x(t))-u(t,x^{(0)}(t))]
dt+\sqrt{\kappa}d w(t)\\
&=& U(t,x(t))dt+\sqrt{\kappa}d w(t)
\eeqn
where $w(t)$ is the standard Brownian motion in $\IR^d$.
It is also useful to consider the associated
backward stochastic flow 
which is the solution of the backward stochastic differential
equation
\beq
\label{bf}
d\Phi^{t}_s(x)&=&-{U}(s,\Phi^{t}_s(x))ds
+
\sqrt{\kappa}dw(t),\quad 0\leq s\leq t\\
\Phi^{t}_t(x)&=&x.
\eeq
Denote by $\IM$ the expectation with respect to
the molecular diffusion and
consider the scalar field $T(t,x)$
\be
\label{1.1'}
T(t,x)\equiv \IM[T_0(\Phi^{t}_0(x))]
\ee
which satisfies
the advection-diffusion equation
\be
\label{ad}
\frac{\partial T(t,x)}{\partial t}= {U}(t,x)\cdot\nabla T(t,x)+
\frac{\kappa}{2}\Delta T(t,x)
,\quad
T(0,x)=T_0(x).
\ee
We interpret eq. (\ref{ad}) in the weak sense
\beq
\lan T(t,\cdot), \theta\ran - \lan T_0, \theta \ran &=&
\frac{\tkappa}{2}\int_0^t \lan T(s,\cdot), \Delta \theta\ran ds
-\int_0^t \lan T(s,\cdot), V(s,\cdot)\cdot\nabla
\theta
\ran ds
\label{weak2}
\eeq
for any test function $\theta \in C^\infty_c(\IR^d)$, the space of
smooth functions with compact supports.

To study the small-scale behavior
we introduce the following scaling limit.
First we assume that the integral and viscous scales of the field $U$ are
$\ell_0=\ep L,
\ell_1=\ep/K$ with $L, K$ tending to $\infty$ in a way to be
specified later. 
Then we re-scale the variables $x\to \ep x, t\to \ep^{2q} t$
amounting to considering
the re-scaled pair separation
\[
x^\ep(t)=\ep^{-1}x(\ep^{2q}t).
\]
The scaling parameter $\ep$ will tend to zero, indicating
that we are considering the emergent
inertial range of scales $\ell_1\ll |x|\ll\ell_0$
(since $K,L\to\infty$) as a result of a large Reynolds number.
We also set
\be
\label{ka}
\kappa=\ep^{2-2q}{\tilde \kappa},\quad\hbox{with}\,\,\tkappa=\tkappa(\ep)
\ee
After re-scaling, the advection-diffusion equation becomes
\be
\frac{\partial T^\ep}{\partial t}=
\ep^{2q-1}{U}(\ep^{2q}t,\ep x)
\cdot\nabla T^\ep+
\frac{\tkappa}{2}\Delta T^\ep.
\label{ad2}
\ee
We take the initial data $T^\ep(0,x)=T_0(x) \in L^\infty(\IR^d)\cap
L^2(\IR^d)$.
 Let 
 \[
 V(t,x)=\ep^{1-\alpha}U(\ep^{2\beta}t, \ep x).
 \]
 As before (cf. (\ref{8''})) the energy spectrum of the
 rescaled field $V$ is given by
     \[
      \cE_{K,L}(\alpha,k)=
        \left\{\begin{array}{ll}
     E_0(\II-k\otimes k|k|^{-2})
       |k|^{1-2\alpha},&\hbox{for}\,\,|k|\in (L^{-1},
          K)\\
          0,& \,\, \hbox{else}
           \end{array}
        \rt.    
      \]
 We 
rewrite eq. (\ref{ad2}) in terms of  $V$ as
\be
\frac{\partial T^\ep}{\partial t}=
\ep^{2q+\alpha-2} V(\ep^{2(q-\beta)}t,x)\cdot\nabla T^\ep+
\frac{\tkappa}{2}\Delta T^\ep, \quad T^\ep(0,x)=T_0(x).
\label{ad3}
\ee
A simple, nontrivial scaling limit is the white-noise limit
when
\be
q<\beta
\label{4b}
\ee
and
\be
\label{4q}
q= 2-\alpha-\beta
\ee
resulting from equating $2q+\alpha-2$ and $q-\beta$.
Inequality (\ref{4b}) and (\ref{4q}) then gives the condition
\be
\alpha+2\beta>2.
\label{low}
\ee
Note that for 
\be
\alpha+\beta<2
\label{up}
\ee
and thus $q>0$ we have  a short-time limit; otherwise, it is
a long time (but small spatial scale) limit.


The paper is organized as follows. In Section~2 we state the
main results and discuss their implications. In Section~3 we discuss
the meaning of solutions for the colored-noise and white-noise
models and prove the uniqueness for the latter.
In Section~4, we prove Theorem~\ref{thm1}:
we prove the tightness of the measures in Section~4.1
and, in Section~4.2, identify the limiting measure by the martingale
formulation.
In Section~5, we prove Theorem~\ref{thm2}.
The method of proof is the same as  that in \cite{ou-kr} (see also
\cite{CF}).
We refer the reader to \cite{Ku}
for the full exposition of the perturbed test
function method used here.
We note that the method of \cite{Kun} requires 
sub-Gaussian behavior and spatial
regularity of the velocity field and is not
applicable here.

\section{Main Theorems and Interpretation}

Let us begin by briefly recalling the Kraichnan model.
The model
  has 
  a white-noise-in-time incompressible
   velocity field
   which can be described as
   the time derivative of a zero mean, isotropic
   Brownian vector field $B_t$ with the  two-time structure function
   \beq
   \nonumber
   \lefteqn{\IE [B_t(x)-B_t(y)]\otimes[B_s(x)-B_s(y)]}\\
   &=&\min{(t,s)}\int 2[1-\cos{(k\cdot (x-y))}]
   a^{-1}\bar{\cE}_L(\eta+1,k)|k|^{1-d
  }dk,\quad\eta\in (0,1)
  \label{4B}
  \eeq
  with
 \[
 \bar{\cE}_L(\eta+1,k)=\lim_{K\to\infty}\cE_{K,L}(\eta+1,k).
 \]
   In this paper, we interpret
   the corresponding advection-diffusion equation 
   for the Kraichnan model in the sense of Stratonovich's integral
   \be
   \label{kr}
   d T_t(x) = \left[\nabla T_t(x)\right]^\dagger
   \circ [d B_t(x)-d B_t(0)]+ \frac{\kappa_0}{2} \Delta T_t(x) \,\,dt,\quad
   \kappa_0\geq 0,\quad T(0,x)=T_0(x)
   \ee
   which can be rewritten as an It\^{o}'s SDE
   \be
   dT_t=
   \left(\frac{\kappa_0}{2}\Delta +
   \frac{1}{a}\bar{\cB}\right)
   T_t\,dt+\sqrt{2}a^{-1/2}\n T_t\cdot d\bar{W}^{(1)}_t
   \label{14.2}
   \ee
   where $\bar{W}_t^{(1)}(x)$ is the Brownian vector field with the spatial
   covariance
   \be
   \bar{\Gamma}^{(1)}(x,y)=\int
   [\exp{(ik\cdot x)}-1][\exp{(-ik\cdot y)}-1]
   \bar{\cE}_L(\eta+1,k)|k|^{1-d}dk,\quad\eta=\alpha+\beta-1
   \label{cov}
   \ee
   and the operator $\bar{\cB}$ is given by
   \beq
   \bar{\cB}\phi(x)&=&\sum_{i,j}\bar{\Gamma}_{ij}^{(1)}(x,x)
   \frac{\partial^2 \phi(x)}{\partial x^i\partial x^j},\quad\phi\in
   C^\infty(\IR^d).
   \label{56}
   \eeq
   We will discuss the meaning of solutions for the Kraichnan model
   and prove the uniqueness property in
   Section~3.
The Kraichnan  model for passive scalar
 has been widely studied to understand turbulent
  transport in the inertial range because of its tractability
  (see, e.g.,  \cite{SS}, \cite{CFKL}, \cite{GK}, \cite{FGV},
  \cite{MK}, \cite{LR}, \cite{GV}, \cite{EV} and the
  references therein).
   The tractability of
   this model lies in the Gaussian and white-noise nature of
   the velocity field.

\begin{theorem}
\label{thm1}
Suppose 
$\alpha+2\beta>2$.
Let
$L<\infty$ be fixed
and let $K=K(\ep)$ such that
$\lim_{\ep\to 0}K=\infty$.
Let $\tkappa=\tkappa(\ep)>0$ such that $\lim_{\ep\to 0}\tkappa=\kappa_0<\infty.$
 Let $T_0\in L^\infty(\IR^d)\cap L^2(\IR^d)$.
 If, additionally,  any one of
 the following conditions is satisfied:
 \begin{itemize}
 \item[(i)] $\alpha+2\beta> 4$;
 \item[(ii)] $\alpha+2\beta=4,\quad \lim_{\ep\to 0}\tkappa\ep^2
 \sqrt{\log{K}}=0$;
 \item[(iii)] $3<\alpha+2\beta<4,\quad\lim_{\ep\to 0}\tkappa\ep^2
 K^{4-\alpha-2\beta}=0$;
 \item[(iv)] $\alpha+2\beta=3,\quad\lim_{\ep\to 0}
 \tkappa\ep^2 K=\lim_{\ep\to 0}\ep
 \sqrt{\log{K}}=0$;
 \item[(v)] $2<\alpha+2\beta< 3,\quad
 \lim_{\ep\to 0}\tkappa\ep^2K^{4-\alpha-2\beta}=
 \lim_{\ep\to 0}\ep K^{3-\alpha-2\beta}=0$
 \end{itemize}
 Then for the exponent $q$ given in (\ref{4q})
  the solution $T_t^\ep$ of (\ref{ad3}) converges in distribution,
  as $\ep\to 0$,
  in the space $D([0,\infty);L^\infty_{w^*}(\IR^d)\cap L^2_w(\IR^d))$
   to the scalar field $T_t$
    for pair dispersion in the  Kraichnan model. 
 The limiting Kraichnan model
 has the spatial covariance given by (\ref{cov}).
 Here $D([0,\infty);L^\infty_{w^*}(\IR^d)\cap L^2_w(\IR^d))$
 is the space of $L^\infty(\IR^d)\cap L^2(\IR^d)$-valued
 right continuous processes with left limits endowed
 with the Skorohod metric \cite{Bi} and $L^\infty_{w^*}(\IR^d)$
 ($L^2_w(\IR^d)$) is
 the standard space $L^\infty(\IR^d)$($L^2(\IR^d)$) 
 endowed with the weak* (weak) topology.
 \end{theorem}
 \begin{remark}
 \label{rmk1}
 In addition to the 
 assumptions stated in the Introduction and in the theorem,
 we use in the proof of Theorems \ref{thm1} the assumption
 \begin{equation}
 \label{1.4'}
 \sup_{t<t_0}\int_{|x|\leq M} |\tilde{V}^\ep_t(x)|\,\,dx
 =o\lt(\frac{1}{\ep}\rt),\quad \ep\to 0, \quad\forall 0<M<\infty
 \end{equation}
 with a random constant possessing
 finite moments where
 \[
 \tilde{V}^\ep_t(x)=
 \frac{1}{\ep^2}\int^\infty_t\IE_{t\ep^{-2}}V(\frac{s}{\ep^2},x)\,ds.
 \]
 For Gaussian velocity fields one
 has
 \begin{equation}
 \nonumber
 M^d\sup_{\substack{|x|\leq M\\t\leq t_0}} \left|\tilde{V}\left(\frac{t}{\vas^2},x\right)\right|\leq
 CL^{\alpha+2\beta-2}\log \left[\frac{M^dt_0}{\vas^2} \right]
 =o\lt(\frac{1}{\ep}\rt)
 \end{equation}
 where the random constant $C$
 has a Gaussian-like tail by Chernoff's bound.
 Condition (\ref{1.4'}) allows certain degree of intermittency in the velocity
 field.
 \end{remark}
 Note that, in Theorem~\ref{thm1}, when $\kappa_0>0$ and $2<\alpha+2\beta<3$,
 $\lim_{\ep\to 0}\tkappa\ep^2K^{4-\alpha-2\beta}=0$ implies 
 $\lim_{\ep\to 0}\ep K^{3-\alpha-2\beta}=0$.
 Also, $\alpha+2\beta<3$ contains the regime $\alpha+\beta<2$
 in which the limiting Brownian velocity field is
 spatially H\"{o}lder continuous
 and has a Hurst exponent $\eta=\alpha+\beta-1\in
 (1/2,1)$, i.e. the limiting velocity field has a {\em persistent}
 spatial correlation.

If we let $L\to\infty$ in the Kraichnan model,
we see that it gives rise to a Brownian velocity field $\bar{B}_t$ 
with the structure function 
\beq
\nonumber
\lefteqn{\IE [\bar{B}_t(x)-\bar{B}_t(y)]\otimes[\bar{B}_s(x)-\bar{B}_s(y)]}\\
&=&
\min{(t,s)}\int 2[1-\cos{(k\cdot (x-y))}]
a^{-1}\bar{\bar{\cE}}(\alpha+\beta,k)
|k|^{1-d}
dk 
\label{17}
\eeq
where
\[
\bar{\bar{\cE}}(\alpha+\beta,k)=\lim_{L\to \infty}\bar{\cE}_L(\alpha
+\beta,k).
\]
The spectral integral in (\ref{17}) is convergent only for 
$\alpha+\beta<2.$
The convergence of the integral in (\ref{17})
means that the limiting Brownian velocity field $\bar{B}_t$
has spatially homogeneous increments.


We can prove the convergence to the Kraichnan model with
velocity field $\bar{B}_t$ in the simultaneous limit
of $\ep\to 0, K, L\to\infty$ if additional conditions
are satisfied:
\begin{theorem}
\label{thm2}
Suppose $\alpha+\beta<2$ and all the assumptions of Theorem~\ref{thm1}
(Thus, only regime {\rm (v)} is relevant) except for
the finiteness of $L$.
Instead, let $L=L(\ep)\to \infty$ such that
\be
\label{3.3}
\lim_{\ep\to 0} L^{2(\alpha+2\beta-2)} \ep =0.
\ee
Then the same convergence holds as in Theorem~\ref{thm1}. The
limiting 
Brownian velocity field $\bar{B}_t$ has the structure function
given by (\ref{17}).
\end{theorem}
\begin{remark}
\label{rmk2}
In addition to the assumptions of Theorem~\ref{thm1} (cf. Remark 1) 
we use in the proof of Theorems \ref{thm2} the assumption
\begin{equation}
\label{1.4''}
\sup_{t<t_0}\int_{|x|\leq M} |\tilde{V}^\ep_t(x)|^2\,\,dx
\leq C L^{2(\alpha+2\beta-2)}\frac{1}{\ep},
\quad \ep\to 0,\,\,L\to\infty, \quad\forall 0<M<\infty
\end{equation}
with a random constant $C$ possessing
finite moments. For Gaussian velocity fields one has 
\[
\sup_{t<t_0}\int_{|x|\leq M} |\tilde{V}^\ep_t(x)|^2\,\,dx
\leq C L^{2(\alpha+2\beta-2)}\lt(\log{\frac{1}{\ep}}\rt)^2,
\quad \ep\to 0,\,\,L\to\infty, \quad\forall 0<M<\infty.
\]
One sees that Condition (\ref{1.4''}) 
is in some sense more tolerant of intermittency than 
(\ref{1.4'}) is.

\end{remark}

Due to the divergence-free property of
the velocity field,
the pre-limit scalar field satisfies the energy identity
(\cite{LU}, Chapt. III, Theorem 7.2)
\be
\label{45}
\int |\tvas(x)|^2 \,dx+\tkappa\int^t_0\int |\n \tvas|^2(x)\,dx\,ds
=\int|T_0(x)|^2\,dx
\ee
provided that $T_0\in L^2(\IR^d)$. From (\ref{45}) we have
the estimates
\[
\|\tvas\|_2^2 <\|T_0\|_2^2,\quad
\int^t_0\|T^\ep_s\|^2_{H^1}\,ds\leq (t+\frac{1}{\tkappa})\|T_0\|^2_2,\quad
t>0
\]
where $\|\cdot\|_{H^1}$ is the norm of the standard Sobolev space
$H^1(\IR^d)$ of square-integrable functions with square-integrable
1st derivative. Thus the law of $T^\ep$ is naturally supported
by the space of continuous $L^2(\IR^d)$-valued processes
which are also in $L_{\hbox{\tiny loc}}^2([0, \infty); H^1(\IR^d))$.
Following \cite{CF} we consider the space
\[
\Omega=D([0,\infty); L^2_w(\IR^d)\cap L^\infty_{w^*} (\IR^d))\cap L^2_{w,
\hbox{\tiny loc}}([0,\infty); H_w^1(\IR^d))
\]
where the subscripts $w$ and $\hbox{ loc}$ denote the weak and
the local topologies, respectively.

In the case of $\tkappa>0,\kappa_0>0$
the above observation and 
the tightness argument for Theorem~\ref{thm1} and \ref{thm2} then  imply 
the tightness of $T^\ep_t$  in the space $\Omega$.
We  have the following corollary.
\begin{cor}
\label{cor1}
If $\kappa_0>0$ and $T_0\in L^\infty(\IR^d)\cap  L^2(\IR^d)$
then the convergence holds in the space $\Omega$ in the
following regimes:
\begin{description}
\item[Case 1] Let $L<\infty$ be fixed and
$K\to \infty$ as $\ep\to 0$. 
\begin{itemize}
\item[(i)]$\alpha+2\beta>4$;
\item[(ii)] $\alpha+2\beta=4, \quad\lim_{\ep\to 0}
\ep^2\sqrt{\log{K}}=0$;
\item[(iii)] $2<\alpha+2\beta<4,\quad \lim_{\ep\to 0}
\ep^2 K^{4-\alpha-2\beta}=0$.
\end{itemize}
\item[Case 2] 
Suppose $\alpha+\beta<2<\alpha+2\beta$ and
$L,K\to \infty $ as $\ep \to 0$ such that
\[
\lim_{\ep\to 0}
\ep^2 K^{4-\alpha-2\beta}=\lim_{\ep\to 0}L^{2(\alpha+2\beta-2)}
\ep=0.
\]
\end{description}
In particular,
\beq
\label{diss}
\lefteqn{\|T_0\|^2_2- \limsup_{\ep\to 0}
\IE[\|T^\ep_t\|_2^2]}\\
\nonumber
&=&\liminf_{\ep\to 0}
\tkappa \int^t_0\IE[\|\nabla T^\ep_s\|_2^2]\,ds\\
&\geq &\kappa_0 
\int^t_0\IE[\|\nabla T_s\|_2^2]\,ds>0
,\quad t>0,\quad\hbox{unless}\,\, T_s\equiv 0,\quad 0\leq s\leq t
\nonumber
\eeq 
where $T_t$ is the solution of the corresponding Kraichnan model.
\end{cor}

In the case of $\tkappa>0,\kappa_0=0$
and $T_0\in L^2\cap L^\infty$,
the limiting Kraichnan model conserves the $L^2$-norm of $T_t$. 
The energy identity
(\ref{45})
then implies
\[
\|\tvas\|_2^2 \leq \|T_0\|_2^2= \|T_t\|_2^2,\quad \forall \ep>0,\,\, \forall t>0
\]
which in turn implies $\lim_{\ep\to 0}\|T^\ep_t\|_2=\|T_t\|_2$.
Hence the weak sense of convergence in Theorem~\ref{thm1} and \ref{thm2} can
be strengthened to the strong $L^2$ convergence.
\begin{cor}
\label{cor2}
If $\kappa_0=0$ and $T_0\in  L^\infty(\IR^d)\cap  L^2(\IR^d)$
then the convergence holds in the space $D([0,\infty);L^2(\IR^d)\cap
L^\infty_{w^*}(\IR^d))$
in the respective regimes listed in Theorem~\ref{thm1} and \ref{thm2}.
In particular,
\[
\|T_0\|^2_2- \lim_{\ep\to 0}
\IE[\|T_t^\ep\|_2^2]=0, \quad t>0\quad\hbox{a.e.}
\]
\end{cor}
We see that in the context of Corollary~1
there is positive dissipation (\ref{diss})
while there is none in the context of Corollary~2.
The conditions of the limit theorems set a constraint
for the presence of positive dissipation:
On the observation scale $\ep$, if the molecular diffusion
$\kappa$ is of order $\ep^{2-2q}$, then there is
always positive dissipation no matter how slow $\ell_1$ vanishes.
On the other hand if $\kappa\ll \ep^{2-2q}$ (i.e.
$\kappa_0=0$) and the dissipation
is positive, then
\[
\ell_1=O(\ep^\nu),\quad\nu=\frac{4-\alpha-2\beta}{3-\alpha-2\beta} 
\]
with $\nu\in (2,\infty)$ in the regime $\alpha+\beta<2<\alpha+2\beta$
(cf. (\ref{range})).
An open question is whether there is a positive dissipation
as $\ep,\kappa\to 0$ with $\ell_1=0$ at the outset. If there is, then
the Kraichnan model (\ref{kr}) is unlikely to be the governing
equation of the scaling limit (if exists).

In the case of $\tkappa=0$, 
a still stronger sense of convergence holds since now eq. (\ref{ad3}) is
of first order and any  locally bounded measurable function $\phi(T^\ep)$
of the scalar field satisfies the same equation (\ref{weak2}) with 
$\tilde\kappa=0$.
The same argument for the proof of Theorem~\ref{thm1} and ~\ref{thm2} 
will then yield the
following result.
\begin{theorem}
\label{thm3}
Assume the conditions stated in Remarks~1 and ~2.
Let $\tkappa=0$, $T_0, \phi(T_0)\in L^\infty(\IR^d)\cap L^2(\IR^d)$ where
$\phi$ is a locally bounded measurable
function
from $\IR$ to $\IR$. 
Then $ \tvas, \phi(\tvas)$ converge in the space
$D([0,\infty);L^\infty_{w^*}(\IR^d)\cap L^2(\IR^d))$ to the corresponding Kraichnan model
in the following regimes.
\begin{description}
\item[Case 1] Let $L<\infty$ be fixed and
$K\to \infty$ as $\ep\to 0$. 
\begin{itemize}
\item[(i)]$\alpha+2\beta>3$;
\item[(ii)] $\alpha+2\beta=3, \quad\lim_{\ep\to 0}
\ep\sqrt{\log{K}}=0$;
\item[(iii)] $2<\alpha+2\beta<3,\quad \lim_{\ep\to 0}
\ep K^{3-\alpha-2\beta}=0$.
\end{itemize}
\item[Case 2]
Suppose $\alpha+\beta<2<\alpha+2\beta$ and
$L,K\to \infty $ as $\ep \to 0$ such that
\[
\lim_{\ep\to 0}
\ep K^{3-\alpha-2\beta}=\lim_{\ep\to 0}L^{2(\alpha+2\beta-2)}
\ep=0.
\]
\end{description}
\end{theorem}
\begin{remark}
\label{rmk3}
The assertions of Theorems \ref{thm1}, \ref{thm2}, \ref{thm3} and
Corollaries \ref{cor1}, \ref{cor2}
hold true for random as
well as deterministic initial data.
\end{remark}

When the parameters are in the regime $\alpha+\beta<2<\alpha+2\beta$,
by taking the expectation in the It\^o's equation with the Brownian
velocity field $\bar{B}_t$ one sees readily that
the
{\em longitudinal} relative diffusion coefficient
is given by
\beq
\label{4.3'}
\frac{\kappa_0}{2}+ \frac{1}{a}\frac{x}{|x|}\cdot\bar{\bar{\Gamma}}^{(1)}(x,x)
\cdot\frac{x}{|x|}
&\approx& \frac{1}{a} C^{-1}_{\alpha+\beta}E_0|x|^{2\eta}\\
&&\hbox{for}\,\,\kappa_0<<1,
\quad \eta=1-q=\alpha+\beta-1\in (1/2,1)
\nonumber
\eeq
with
\beq
\label{4.3}
\nonumber
\bar{\bar{\Gamma}}^{(1)}(x,x)&=&\lim_{L\to\infty}{\bar{\Gamma}}^{(1)}(x,x)
=C_{\alpha+\beta}^{-1} E_0 |x|^{2\eta}\left[(1+\frac{2(\alpha+\beta-1)}{d-1})\II
-\frac{2(\alpha+\beta-1)}{d-1}x\otimes  x|x|^{-2}\rt]
\eeq
where $C_{\alpha+\beta}$ is defined as in (\ref{cd}),
except with $\alpha$ replaced by $\alpha+\beta$.
The exponent $q$ is related to the exponent $p$
in the expression for the mean square pair separation
as follows:
\be
\label{t3}
\IE|x|^2(t)\sim a^{-p}E_0^p t^p,\quad p=1/q=1/(2-\alpha-\beta)
\ee
up to a dimensionless constant depending only on $\alpha+\beta$.
Expressions (\ref{t3}) and (\ref{4.3'}) 
can be viewed as the generalization of Richardson's $t^3$ and
$4/3$-laws, respectively.
In general,
$p\in (2,\infty)$, indicating super-ballistic (i.e. accelerating) motion
as a result of a scale-dependent relative diffusivity.

We now remark on the range of scales for which Theorem~\ref{thm2} is proved
and Richardson's laws can be reasonably interpreted. 
Let $\ep$ be the scale of dispersion.
Then the limit theorem holds in the range
\be
\label{gamma}
\ep\ll \min{\lt[ \ell_1^{\gamma},
\lt(\frac{1}{\ell_0}\rt)^{\frac{2(\alpha+2\beta-2)}{5-2\alpha-4\beta}}\rt]},
\quad \gamma=\left\{
\begin{array}{ll}
\frac{3-\alpha-2\beta}{4-\alpha-2\beta},&\hbox{if}\,\,\kappa_0=0\\
\frac{4-\alpha-2\beta}{6-\alpha-2\beta},&\hbox{if}\,\,\kappa_0>0
\end{array}
\right.
\ee
In the usual situation with $\ell_0=O(1)$ the range of scales
covered by the limit theorem  has an upper limit of
\be
\label{range}
\ell_1^{\gamma}, 
\quad \hbox{with}\quad\gamma\in
\left\{\begin{array}{ll}
(0,1/2),&\hbox{if}\,\,\kappa_0=0\\
(1/3,1/2), &\hbox{if}\,\,\kappa_0>0
\end{array}
\right.
\quad\hbox{for}\quad \alpha+2\beta>2>\alpha+\beta,
\ee
which is limited to the low end of
the inertial range depending on $\alpha,\beta, \kappa_0$.
It is not clear whether this is physical or a technical matter.
Qualitatively similar restriction of Richardson's laws in
synthetic flows has been observed in
numerical calculation (cf. \cite{BCCV}, \cite{FV}).

If we stretch the validity of (\ref{t3}) and (\ref{4.3'})
by taking the limit $\alpha\to 4/3, \beta\to 1/3$ from within the
valid regime, 
the resulting  exponents are
$p=3, 2\eta=4/3$ in accordance with Richardson's laws. 
On the boundary  $\alpha+2\beta=2$ the scaling exponent $q$ should be given by
\be
\label{r3}
q=\beta=1-\alpha/2
\ee
which also coincides with the limiting value of (\ref{4q}). 
With (\ref{r3}) and $K,L\to\infty$,
the solution of (\ref{ad3}) converges to that
of the advection-diffusion equation with the molecular
diffusivity $\kappa_0=\lim_{\ep\to 0}\tkappa$ and
the time-stationary,
spatially H\"{o}lder continuous velocity
field $\bar{V}$ whose two-time correlation function is
\beqn
\lefteqn{\IE\lt[ \bar{V}(t,x)\otimes \bar{V}(s,y)\rt]}\\
&=&\int_{\IR^d}[\exp{(i k\cdot x)}-1][\exp{(-i k\cdot y)}-1]
\exp{(-a|k|^{2-\alpha}|t-s|)}
\bar{\bar{\cE}}(\alpha, k) |k|^{1-d}d k,\alpha\in (1,2)
\nonumber
\eeqn
which has the self-similar structure
\[
\IE\lt[ \bar{V}(\lambda^{2\beta}t,\lambda x)\otimes \bar{V}(\lambda^{2\beta}s,
\lambda y)\rt]=\lambda^{2\alpha-2}\IE\lt[ \bar{V}(t,x)\otimes \bar{V}(s,y)\rt].
\]
In view of the 4th order scale-invariance property
it is reasonable to postulate the temporal
self-similarity 
on the mean-square
relative dispersion as $\kappa_0\to 0$
\[
\IE|x(t)|^2=f(E_0,a) t^{1/\beta},
\]
which has the same exponent 
as the limiting case of (\ref{t3}) as $\alpha+2\beta\to 2$,
where the unknown function $f$ satisfies the relation
\[
f(E_0,\lambda a)\lambda^{-1/\beta}=f(\lambda^{-2}E_0,a),\quad\forall\lambda>0.
\]
Dimensional analysis with (\ref{const22}) then leads to the relation
\[
\IE|x(t)|^2=\bar{C}_R C_\alpha^{-1/(2\beta)} E_0^{1/(2\beta)}t^{1/\beta},
\]
where $\bar{C}_R$ is the generalized Richardson constant.
For $\beta=1/3$ the exponent $p$ is $1/3$ as predicted by
Richardson's $t^3$-law. However, since the limiting velocity field
is non-white-in-time, the notion of relative
diffusivity is not strictly well-defined.
Therefore the temporal memory persists on small or intermediate time scales
and the notion of relative diffusivity does not describe accurately
the process of relative dispersion on the boundary $\alpha+2\beta=2$.
(cf., e.g., \cite{HP}, \cite{FV} and
\cite{MK}).

Let us consider the regime $\alpha+2\beta<2$. The correct scaling is
to set
\be
\label{q3}
2q+\alpha-2=0\quad\hbox{or}\quad q=1-\alpha/2.
\ee
Then the exponent $2(q-\beta)$ of the temporal scaling in (\ref{ad3})
is positive due to $\alpha+2\beta<2$, meaning the time variable
is slowed down as $\ep\to 0$. It is easy to see by
a regular perturbation argument that the solution $\tvas$ converges
in the sense described in Theorem~\ref{thm1} to the solution $\bar{T}_t$
of the following equation
\[
\frac{\partial\bar{T}_t}{\partial t}=V(0,x)\cdot\nabla \bar{T}_t+
\frac{\kappa_0}{2}
\Delta \bar{T}_t,\quad \bar{T}_0=T_0\in L^\infty(\IR^d)
\]
if $\kappa_0>0$. If, however, $\kappa_0=0$, the above equation
probably have multiple solutions for a given initial condition. 
The relation (\ref{q3}) is consistent with
the numerical simulation  using two-dimensional frozen velocity fields
with Kolmogorov-type spectrum \cite{EM}. 

Unlike the previous regime,
for either $\alpha+2\beta=2$ or $\alpha+2\beta<2$ there is no
restriction on the vanishing rate of $\ell_1$.
\section{Formulation}

From the general theory of parabolic partial differential equations
\cite{Fr}, for any fixed $\tkappa>0, \ep>0$, 
there is a unique $C^{2+\eta}$-solution $T^\ep_t(x)$, 
$0<\forall \eta<\alpha-1$.
But the solutions $T^\ep_t$ may lose all the regularity
as  $\tkappa\to 0, \ep\to 0$.
So we consider the
weak formulation  of the equation:
\beq
\lan T_t^\varepsilon, \theta\ran - \lan T_0, \theta \ran &=&
\frac{\tkappa}{2}\int_0^t \lan T_s^\vas, \Delta \theta\ran ds
-\frac{1}{\vas}\int_0^t \lan T_s^\vas, V(\frac{s}{\vas^2},\cdot)\cdot
\nabla\theta
\ran ds
\label{weak}
\eeq
for any test function $\theta \in C_c^\infty(\IR^d)$,
the space of smooth functions with compact support.
On the other hand the energy identity (\ref{45}) implies
 $\tvas\in L^2([0,t_0];H^1(\IR^d))$ if $T_0\in L^2(\IR^d)$. 
 Hence for $L^2$ initial data the prelimit
measure $\IP^\ep$ is supported in the space $L^2([0,t_0];H^1(\IR^d))$
and, by the tightness result (Section~4.1),
the limiting measure $\IP$ is supported
in $L_w^2([0,t_0];H_w^1(\IR^d))$.

As in (\ref{bf}) and (\ref{1.1'}) the solutions $\tvas$ can be represented
as
\be
\label{1.1}
\tvas =\IM[T_0(\Phi^{t,\ep}_0(x))]
\ee
where $\Phi^{t,\ep}_s(x)$ is the unique stochastic flow
satisfying
\beq
\label{back-flow}
d\Phi^{t,\ep}_s(x)&=&-\frac{1}{\ep}{V}(\frac{s}{\ep^2},\Phi^{t,\ep}_s(x))ds
+
\sqrt{\tkappa}dw(t),\quad 0\leq s\leq t\\
\Phi^{t,\ep}_t(x)&=&x.
\eeq
In the case of $\tkappa=0$, $\Phi^{t,\ep}_0(x), \forall t,$ is
almost surely a diffeomorphism of $\IR^d$ and
$\tvas =T_0(\Phi^{t,\ep}_0(x))$. Moreover,
for any locally bounded measurable function $\phi:\IR\to\IR$, 
$\phi(T^\ep_t(x))=(\phi\circ T_0)(\Phi^{t,\ep}_0(x))$.

In view of the averaging in the representation (\ref{1.1}) we have
\begin{prop}\label{prop:1}
\[
\|T^\ep_t\|_\infty \leq \|T_0\|_\infty \quad\hbox{a.s.}
\]
\end{prop}
Clearly, Proposition~1 holds for the case of $\tkappa=0$ as well.

For tightness as well as identification of the limit,
 the following infinitesimal operator  $\cA^\ep$ will play an important role.
Let $V^\vas_t\equiv V(t/\ep^2,\cdot)$. Let $\mathcal{F}_t^\vas$ be the
$\sigma$-algebras generated by $\{V_s^\vas, \, s\leq t\}$  and
$\mathbb{E}_t^\vas$ the corresponding conditional expectation w.r.t. $\cF^\ep_t$.
Let $\cM^\ep$ be the space of measurable function adapted to $\{\cF^\ep_t, \forall t\}$ 
such that $\sup_{t<t_0}\IE|f(t)|<\infty$.
We say $f(\cdot)\in \cD(\cA^\ep)$, the domain of $\cA^\ep$, and $\cA^\ep f=g$
if $f,g\in \cM^\ep$ and for
$f^\delta(t)\equiv\delta^{-1}[\IE^\ep_t f(t+\delta)-f(t)]$
we have
\beqn
\sup_{t,\delta}\IE|f^\delta(t)|&<&\infty\\
\lim_{\delta\to 0}\IE|f^\delta(t)-g(t)|&=&0,\quad\forall t.
\eeqn
 For $f(t)=\phi(\lan T_t^\vas, \theta\ran),
 f'(t)=\phi'(\lan T_t^\vas, \theta\ran),
 \forall \phi\in C^\infty(\IR)$
we have the following expression
from (\ref{weak}) and the chain rule 
\beq
\label{gen}
 \cA^\vas
f(t)
&=& \frac{\tkappa}{2} f'(t) \lan T_t^\vas,
\Delta \theta\ran - \frac{1}{\vas} f'(t) \lan T_t^\vas,
\cv(\theta)\ran
\eeq
where 
\be
\label{nu2}
\cv(\theta) \equiv V^\ep_t \cdot\n\theta.
\ee
 A main property of $\cA^\ep$ is
that 
\be
\label{12.2}
f(t)-\int^t_0 \cA^\ep f(s) ds\quad\hbox{is a  $\cF^\ep_t$-martingale},
\quad\forall f\in \cD(\cA^\ep).
\ee
Also,
\be
\label{mart}
\IE^\ep_s f(t)-f(s)=
\int^t_0 \IE^\ep_s \cA^\ep f(\tau) d\tau\quad \forall s<t \quad\hbox{a.s.}
\ee
(see \cite{Kur}).

Likewise we formulate the solutions for the Kraichnan model
(\ref{14.2}) as the solutions to the corresponding martingale problem:
Find a measure $\IP$ (of $T_t$) on
the space $D([0,\infty);L^\infty_{w^*}(\IR^d))$
such that
 \beq
 \label{13.2}
 &&f( \lan T_t,\theta\ran)-\int_0^t
 \bigg\{f'( \lan T_s,\theta\ran)\left[\frac{\kappa_0}{2}\lan
 T_s,\Delta\theta\ran+\frac{1}{a}\lan T_s,\bar{\cB}^*\theta\ran\right]
  + \frac{1}{a}f''(\lan T_s,\theta\ran) 
  \lan\theta, \bar{\cK}^{(1)}_{T_s}\theta\ran\bigg\}\,ds\\
\nonumber&&\hbox{{\em is a martingale w.r.t. the filtration of a cylindrical
Wiener process, for each} $f\in C^\infty(\IR)$}
\eeq
where $\bar{\cB}^*$ is the adjoint of $\bar{\cB}$ and
\be
\label{r10}
 \lan\theta, \bar{\cK}^{(1)}_{T_s}\theta\ran
 =\int\int T_s(x)T_s(y)\nabla \theta(x)\cdot\bar{\Gamma}^{(1)}(x,y)\cdot
 \n \theta(y)
 \,dy
 \ee
with $\bar{\Gamma}^{(1)}(x,y)$ given, respectively, by (\ref{cov})
and
 \beq
 \bar{\Gamma}^{(1)}(x,y)&=&\int
 [\exp{(ik\cdot x)}-1][\exp{(-ik\cdot y)}-1]
 \bar{\bar{\cE}}(\eta+1,k)|k|^{1-d}dk,\quad\eta=\alpha+\beta-1
 \label{cov2}
 \eeq
for $L<\infty$ and $L=\infty$.
To identify the limit for the proof of convergence one needs
the uniqueness of solution to the martingale problem (\ref{13.2})
which can be easily obtained as follows.

Taking expectation of (\ref{13.2}) with
$f(r)=r^n, n\in\IN$
we get
for the $n-$point correlation function
\[
F^t_n(x_1,x_2,x_3,...,x_n)\equiv\IE_{T_0}\lt[T_t(x_1)T_t(x_2)\cdots T_t(x_n)\rt]\]
the equation
\beqn
\lefteqn{\lan F^t_n,\otimes^n \theta\ran
-\lan F^0_n,\otimes^n \theta\ran}\\
&=&\int^t_0\left[\sum_{j}\frac{\kappa_0}{2}\lan F^s_n,\theta(x_1)\cdots\Delta\theta(x_j)\cdots\theta(x_n)\ran+\sum_{j}\frac{1}{a}\lan
F^s_n,\theta(x_1)\cdots\bar{B}^*\theta(x_j)\cdots \theta(x_n)\ran\right.\\
&&\left.+\sum_{i<j}\frac{2}{a}\lan F^s_n,\bar{\Gamma}^{(1)}(x_i,x_j):\theta(x_1)\cdots
\nabla\theta(x_i)\cdots\nabla\theta(x_j)\cdots\theta(x_n)\ran\right]\,\,ds
\eeqn
which induces a weakly continuous (hence strongly continuous)
sub-Markovian semigroup on $L^p(\IR^{nd})$,
$\forall p\in (1,\infty)$.
The sub-Markovianity property is inherited from the pre-limit process
$T^\ep_t$.
The generator of the semigroup is given formally as
\be
\label{ndiff}
 \cL_n \Phi(x_1,\cdots,x_n)\equiv \frac{\kappa_0}{2}\sum_{j=1}^n
\Delta_{x_j} \Phi+\frac{1}{a}\sum_{i,j=1}^n \bar{\Gamma}^{(1)}(x_i,x_j):
\nabla_{x_i}\nabla_{x_j}\Phi,\quad\Phi\in C^\infty_c(\IR^{nd}),\,\,
\kappa_0\geq 0
\ee
with the spatial covariance tensor $\bar{\Gamma}^{(1)}(x_i,x_j)$
given by (\ref{cov}) and (\ref{cov2}), respectively, for
$L<\infty$ and $L=\infty$.
Note that the symmetric operator $\cL_n$ (\ref{ndiff})
is an essentially self-adjoint positive operator  on $C_c^\infty(\IR^N), N=nd$
which then induces a {\em unique} symmetric Markov semigroup 
of contractions on $L^2(\IR^N)$.
The essential self-adjointness
is due to the sub-Lipschitz growth of the square-root
of $\bar{\Gamma}^{(1)}(x_1, x_2)$ at large $|x_1|, |x_2|$
(hence no escape to infinity) \cite{Da}.

By Theorem~1.4.1 of \cite{Da2} this semigroup induces a sub-Markovian
$C_0$-semigroup on $L^p(\IR^N)$, $p\in [1,\infty)$. 
The uniqueness holds for
these semigroups in their respective space as well
but we will not pursue it here.

\commentout{
We now show  the uniqueness of such a
semigroup on $L^1(\IR^N)$,
by using the general result of \cite{St} (see also
\cite{Ha}). 
\begin{prop}\cite{St}
The following statements are equivalent:
\begin{itemize}
\item[(i)] There is a unique $C_0$-semigroup on  $L^1(\IR^N)$
whose generator
extends $(\cL_n,C^\infty_c(\IR^N))$.

\item[(ii)] The closure $\bar{\cL_n}$ on $L^1(\IR^{N})$ 
of $(\cL_n,C^\infty_c(\IR^N))$ generates a $C_0$-semigroup.

\item[(iii)] 
The $C_0$-semigroup $\cP_t$, generated by
the Friedrichs' extension of $(\cL_n,C^\infty_c(\IR^N)))$,
is conservative, namely,
its dual semigroup $\cP^*_t$ on $L^\infty(\IR^d)$  satisfies
\[
\cP^*_t 1=1,\quad \forall t>0.
\]
\end{itemize}
\end{prop}

\begin{prop}
\label{SV}
\cite{SV}
If $w(x)$ is a strictly positive $C^\infty$ function on $\IR^N$ with
$w(x)\to +\infty$ as $x\to \infty$ and if $\cL_n w(x)\leq \lambda w(x)$ for
some $\lambda>0$ and all $x$ outside some compact subset
then $\cR 1=1$ where $\cR=(1-\cL_n)^{-1}$ and hence
the associated semigroup is conservative.
\end{prop}
We verify the condition of Proposition~\ref{SV} by considering
any strictly positive, $C^\infty$ radial function 
$w=w(r), r=\sqrt{\sum_j|x_j|^2}$
which equals $\log{r}$ for sufficiently large $r$.
The conditions are satisfied again due to
the sub-Lipschitz growth of the square-root
of $\bar{\Gamma}^{(1)}(x_1, x_2)$ at large $|x_1|, |x_2|$.

}

\section{Proof of Theorem~\ref{thm1}}
\subsection{Tightness}

In the sequel we will adopt the following notation 
\[
f(t)\equiv f(\lan T_t^\vas, \theta\ran),\quad
 f'(t)\equiv f'(\lan T_t^\vas, \theta\ran),\quad
f''(t)\equiv f''(\lan T_t^\vas, \theta\ran),\quad 
\quad\forall f\in C^\infty(\IR).
\]
Namely, the prime stands for the differentiation w.r.t. the original argument (not $t$)
of $f, f'$ etc.

A family 
of processes $\{T^\ep, 0<\ep<1\} \subset D([0,\infty);L^\infty_{w^*}(\IR^d)) 
$ is 
tight if and only if the family of 
processes
$\{\lan T^\ep, \theta\ran, 0<\ep <1\}
\subset D([0,\infty);L^\infty_{w^*}(\IR^d))
$ is tight for all $\theta\in C^\infty_c(\IR^d)$. 
We use the tightness  criterion of \cite{Ku}
(Chap. 3, Theorem 4), namely, 
we will prove:
Firstly,
\be
\label{trunc}
\lim_{N\to \infty}\limsup_{\ep\to 0}\IP\{\sup_{t<t_0}|\lan T_t^\ep, \theta\ran|
\geq N\}=0,\quad\forall t_0<\infty.
\ee
Secondly, for  each $f\in C^\infty(\IR)$  
there is a sequence
$f^\ep(t)\in\cD(\cA^\ep)$ such that for each $t_0<\infty$
$\{\cA^\ep f^\ep (t), 0<\ep<1,0<t<t_0\}$ 
is uniformly integrable and
\be
\label{19}
\lim_{\ep\to 0} \IP\{\sup_{t<t_0} |f^\ep(t)-
f(\lan T^\ep, \theta\ran) |\geq \delta\}=0,\quad \forall \delta>0.
\ee
Then it follows that the laws of
$\{\lan T^\ep, \theta\ran, 0<\ep <1\}$ are tight in the space
of $D([0,\infty);L^\infty_{w^*}(\IR^d))$

Condition (\ref{trunc}) is satisfied as a result of Proposition~1. 
Let
\[
f_1^\vas (t)\equiv \frac{1}{\vas}\int_t^\infty
\mathbb{E}_t^\vas\, f'(t) \lan T_t^\vas,\cV^\ep_s(\theta)\ran\,ds
\]
be the 1-st perturbation of $f(t)$. 
We obtain
\begin{equation}
\label{1st}
f_1^\vas (t)= 
\frac{\vas}{a}
f'(t) \lan T_t^\vas,\cvtil(\theta)\ran
\end{equation}
with
\bea
\label{cvtil}
\cvtil(\theta) &=&\tilde{V}^\ep_t\cdot\n\theta\\
\tilde{V}_t^\vas 
&\equiv&\tilde{V} \left(\frac{t}{\vas^2}, \cdot\right)\equiv\frac{1}{\ep^2}
\int_t^\infty
\IE_t^\vas\, V_s^\vas \, ds
\eea
where $\tilde{V}$ has the power spectrum 
$\cE_{K,L}(\alpha+2\beta,k)$ by  the spectral representation
\begin{equation}
\begin{split}
\label{cond-exp}
\mathbb{E}_t^\vas\, V_s^\vas = \int [e^{ix\cdot k}-1]
e^{-a |k|^{2\beta}|s-t|\vas^{-2}} \widehat{V}_t^\vas (dk),
 \quad \forall \, s\geq t.
 \end{split}
 \end{equation}
Note that while $\vep$ loses differentiability
as $K\to \infty$,  $\veptil$ is almost surely a $C^{1,\eta}$-function
in the limit with
\[
0<\forall \eta<\alpha+2\beta-2
\]
and has uniformly bounded local $W^{1,p}$-norm, $p\geq 1$.

\begin{prop}\label{prop:2}
\begin{enumerate}
$$\lim_{\ep\to 0}\sup_{t<t_0} \mathbb{E} |f_1^\vas(t)|=0,\quad
\lim_{\ep\to 0}\sup_{t<t_0} |f_1^\vas(t)|= 0
\quad \hbox{in probability}$$.
\end{enumerate}
\end{prop}

\begin{proof}
By Proposition \ref{prop:1} we have
\be
\label{1.2}
\mathbb{E}[|f_1^\vas(t)|]\leq \frac{\vas}{a} \|f'\|_\infty
\|T_0\|_\infty 
\|\theta\|_\infty\int_{|x|\leq M}\IE|\tilde{V}^\ep_t|dx
\ee
and
\be
\label{1.3}
\sup_{t< t_0} |f_1^\vas(t)|
 \leq \frac{\vas}{a}
\|f'\|_\infty  \|T_0\|_\infty
\|\theta\|_\infty\sup_{t<t_0}\int_{|x|\leq M}|\tilde{V}^\ep_t|dx.
\ee
By the temporal stationarity of $\tilde{V}^\ep_t$ we can replace the
term
$\IE|\tilde{V}^\ep_t(x)|$ in (\ref{1.2}) by 
$\IE|\tilde{V}(0,x)|$.
By assumption (cf. (\ref{1.4'}), Remark~1),
we have the desired estimate.
Proposition~\ref{prop:2}
now follows from (\ref{1.2}), (\ref{1.3}) and (\ref{1.4'}).
\end{proof}

Set $f^\ep(t)=f(t)-f^\ep_1(t)$.
A straightforward calculation yields
\begin{align*}
\begin{split}
 \cA^\vas f_1^\vas &=-\frac{\tilde{\kappa}\vas}{2a}f''(t)\lan
T_t^\vas,\Delta\theta \ran\lan T_t^\vas,
\cvtil(\theta) \ran +\frac{\tilde{\kappa}\vas}{2a}f'(t)\lan
 T_t^\vas,\Delta\cvtil(\theta)\ran\\
&\quad +
\frac{1}{a}f''(t)\lan T_t^\vas,\cv(\theta)\ran\lan
T_t^\vas,\cvtil(\theta)\ran  -
\frac{1}{a} f'(t)\lan\tvas,\cv(\cvtil(\theta))\ran \\
&\quad -
\frac{1}{\vas}f'(t)\lan\tvas,\cv(\theta)\ran
\end{split}
\end{align*}
and, hence
\begin{align}
\label{25'}
\begin{split}
\cA^\vas f^\ep(t)
&=\frac{\tkappa}{2}f'(t)\lan \tvas,\Delta\theta \ran + \frac{1}{a}f'(t)\lan\tvas,
\cv(\cvtil(\theta))\ran
-\frac{1}{a}f''(t)\lan\tvas, \cv(\theta)\ran\lan\tvas,
\cvtil(\theta)\ran \\
&\quad +
\frac{\tkappa\vas}{2a}\left[f''(t)\lan\tvas,\Delta\theta\ran\lan\tvas,
\cv(\theta)\ran
 -f'(t)\lan\tvas,
\Delta\cvtil(\theta)\ran\right] \\
&=A_1^\vas(t)+A_2^\vas(t)+A_3^\vas(t)+A_4^\vas(t)
\end{split}
\end{align}
where $A_2^\vas(t)$ and $A_3^\vas(t)$ are the $O(1)$ statistical coupling
terms.

For the tightness criterion stated in the beginnings of the section,
it remains to show
\begin{prop}
$\{\cA^\ep f^\ep\}$ are uniformly integrable and
$$\lim_{\ep\to 0}\sup_{t<t_0}\IE|A^\ep_4(t)|=0$$.
\end{prop}

\begin{proof}
We show that $\{A^\ep_i\}, i=1,2,3,4$ are uniformly integrable. 
To see this, we have the following estimates.
\bean
|A_1^\vas(t)|=\frac{\tkappa}{2}\lt|f'(t)\lan \tvas,\Delta\theta \ran\rt|
\leq\frac{\tkappa}{2}\|f'\|_\infty\|T_0\|_\infty\|\Delta \theta\|_1
\eean
Thus $A^\ep_1$ is uniformly integrable since it is uniformly bounded.
\beqn
|A_2^\vas(t)|&=&\frac{1}{a}\lt|f'(t)\lan\tvas, 
\cv(\cvtil(\theta))\ran\rt|\\
&\leq& \frac{C}{a}\|f'\|_\infty\|T_0\|_\infty
\left[
\int_{|x|<M}|{V}^\ep_t|^2\,dx
\right]^{1/2}
\left[\int_{|x|<M}|\nabla\tilde{V}^\ep_t|^2\,dx
\right]^{1/2}.
\eeqn
Similarly,
\beqn
|A_3^\vas(t)|&=& \frac{1}{a}\lt|f''(t)\lan\tvas,\cv(\theta)\ran\lan\tvas,
\cvtil(\theta)\ran\rt|\\
&\leq& \frac{C}{a}\|f''\|_\infty\|T_0\|_\infty^2
\left[\int_{|x|<M}|{V}^\ep_t|^2\,dx+
\int_{|x|<M}|\tilde{V}^\ep_t|^2\,dx
\right].
\eeqn
Thus $A^\ep_2$ and $A^\ep_3$ are uniformly integrable in view of
the uniform boundedness of the 4-th moment of
$V^\ep_t, \tilde{V}^\ep_t$ and $\nabla\tilde{V}^\ep_t$
as $L<\infty$ is fixed and $K\to\infty$ (the 4th order scale-invariance).
 \beq
 |A^\ep_4|
  \nonumber
 &=&\frac{\tkappa\vas}{2a}
 |f''(t)\lan\tvas,\Delta\theta\ran\lan\tvas,
 \cvtil(\theta)\ran
  - f'(t)\lan\tvas,\Delta\cvtil(\theta)\ran\\
  \nonumber
  &\leq& \frac{C\tkappa\vas}{2a}
  \left[\|f''\|_\infty\|T_0\|_\infty^2
\left[\int_{|x|<M}|\tilde{V}^\ep_t|^2\,dx
\right]^{1/2}\right.
 +
\|f'\|_\infty\|T_0\|_\infty \times\\ &&\quad\left.\left[
\int_{|x|<M}|\tilde{V}^\ep_t|^2\,dx+
\int_{|x|<M}|\n\tilde{V}^\ep_t|^2\,dx
+\int_{|x|<M}|\Delta\tilde{V}^\ep_t|^2\,dx
\right]^{1/2}
\right].
 \label{1.10}
  \eeq
  The most severe term in the above argument as a result of $K\to\infty$
  is
  \[
  \frac{\tkappa\ep}{2a}\lt|f'(t)\lan
  \tvas,\Delta\cvtil(\theta)\ran\rt|
  \]
  whose second moment can be bounded as
  \beq
  \nonumber
  \lefteqn{\frac{\tkappa\ep}{2a}\sqrt{\IE\lt|f'(t)\lan
  \tvas,\Delta\cvtil(\theta)\ran\rt|^2}}\\
  \nonumber
  &\leq&C_1\frac{\tkappa\ep}{2a}\|f'\|_\infty
  \|T_0\|_\infty\lt(\int_{|x|<M}\IE\lt[|\Delta\veptil|^2\rt]\,dx\rt)^{1/2}
  \\
  &\leq& C_2\tkappa\ep\times
  \left\{\begin{array}{ll}
  K^{3-\alpha- 2\beta},&\hbox{for}\,\,\alpha+2\beta <3\\
  \sqrt{\log{K}},&\hbox{for}\,\,\alpha+2\beta=3\\
  1,&\hbox{for}\,\, \alpha+2\beta>3
  \end{array}\rt.
  \label{39}
  \eeq
  and, thus, vanishes in the limit by the assumptions of the theorem.
  The 4-th moment behaves the same way by the 4th order scale-invariance.
   Hence
  $A^\ep_4$ is uniformly integrable.
Clearly $$\lim_{\ep\to 0}\sup_{t<t_0}\IE|A^\ep_4(t)|=0.$$
  \end{proof}

\subsection{Identification of the limit}
Once the tightness is established we can use another result 
 in \cite{Ku} (Chapter 3, Theorem 2) to identify the limit.
Let $\overline{\cA}$ be a diffusion or jump diffusion operator such that
there is a unique solution $\omega_t$ in the space 
$D([0,\infty);L^\infty_{w^*}(\IR^d))$
 such that 
\be
\label{38}
f(\omega_t)-\int^t_0\overline{\cA} f(\omega_s)\,ds
\ee
is a martingale. 
We shall show that for each $f\in C^\infty(\IR)$ there exists $f^\ep\in \cD({\cA}^\ep)$
such that
\beq
\label{38.2}
\sup_{t<t_0,\ep}\IE|f^\ep(t)-f(\lan T^\ep_t,\theta\ran)|&<&\infty\\
\label{39.2}
\lim_{\ep\to 0}\IE|f^\ep(t)-f(\lan T^\ep_t,\theta\ran)|&=&0,\quad \forall t<t_0\\
\label{40.2}
\sup_{t<t_0,\ep}\IE|{\cA}^\ep f^\ep(t)-\overline{\cA} f(\lan T^\ep_t,\theta\ran)|&<&\infty\\
\lim_{\ep\to 0}\IE|{\cA}^\ep f^\ep(t)-\overline{\cA} f(\lan T^\ep_t,\theta\ran)|&=&0,\quad
\forall t<t_0.
\label{42.2}
\eeq
Then the aforementioned theorem implies that any tight processes
$\lan T^\ep_t,\theta\ran$ converge
in law to the unique process generated by $\overline{\cA} $.
As before we adopt the notation $f(t)=f(\lan T^\ep_t,\theta\ran)$.

For this purpose,
we introduce the next perturbations $f_2^\ep, f_3^\ep$.
Let
\bea
\label{40.3}
A_2^{(1)}(\phi) &\equiv&\lan\theta,\cK^{(1)}_\phi\theta\ran\\
A_3^{(1)}(\phi)&\equiv&\lan\phi, \IE\lt[\cv(\cvtil(\theta))\rt]\ran
\eea
where the positive-definite operator $\cK^{(1)}_{\phi} $ is defined
as
\beq
\label{15.2'}
\cK^{(1)}_{\phi}\theta&=&
\int\theta(y) \nabla \phi(x)\cdot \Gamma^{(1)}(x,y)\n \phi(y)
\,dy\\
    {\Gamma}^{(1)}(x,y)&=&\int
       [\exp{(ik\cdot x)}-1][\exp{(-ik\cdot y)}-1]
          {\cE_{K,L}}(\alpha+\beta,k)|k|^{1-d}dk
\eeq
such that
\beq
\lan\theta_1,\cK^{(1)}_{T_t}\theta_2\ran& =&
\iint\phi(x)\phi(y)G^{(1)}_{\theta_1,\theta_2}(x,y)\,dx\,dy\\
G^{(1)}_{\theta_{1},\theta_{2}}&\equiv&
\sum_{i,j}\frac{\partial^2}{\partial
x^i
\partial
y^j}\left[\theta_1(x)\theta_2(y)\Gamma_{ij}^{(1)}(x,y)\right]
\eeq
(cf. \ref{r10}).

It is easy to see that
\begin{align}
A_2^{(1)}(\phi)&=\mathbb{E}\lt[\lan\phi, \cv(\theta)\ran
\lan\phi, \cvtil(\theta)\ran\rt]\\
A_3^{(1)}(\phi)&= \lan\cB\phi,\theta\ran
\label{41.2}
\end{align}
where the operator $\cB$ is given by
\[
{\cB}\phi(x)=\sum_{i,j}{\Gamma}_{ij}^{(1)}(x,x)
   \frac{\partial^2 \phi(x)}{\partial x^i\partial x^j}.
   \]

Define
\begin{align*}
f_2^\vas(t) &\equiv
\frac{1}{a}f''(t)\int_t^\infty \mathbb{E}_t^\vas
\lt[\lan\tvas, \cV^\ep_s(\theta)\ran\lan\tvas,\tilde{\cV}^\ep_s(\theta)\ran -A^{(1)}_2(\tvas)\rt]\,ds
\\
f_3^\vas(t) &\equiv \frac{1}{a}f'(t)\int_t^\infty \mathbb{E}_t^\vas
\lt[\lan\tvas, \cV^\ep_s(\tilde{\cV}^\ep_s(\theta))\ran-A^{(1)}_3(\tvas)\rt]\,ds.
\end{align*}
Let
\begin{align*}
G^{(2)}_{\theta_{1},\theta_{2}}(x,y)&\equiv\sum_{i,j}
{\Gamma}_{ij}^{(2)}(x,y)
\frac{\partial \theta_1(x)}{\partial x^i}
\frac{\partial \theta_2(y)}{\partial y^j}\\
\lan\theta_1,\cK^{(2)}_\phi \theta_2\ran &\equiv
\iint\phi(x)\phi(y)G^{(2)}_{\theta_1,\theta_2}(x,y)\,dx\,dy
\end{align*}
where the covariance function $\Gamma^{(2)}(x,y)\equiv
\IE\lt[\veptil(x)\otimes\veptil(y)\rt]$ has the spectral density
$\cE_{K,L}(\alpha+2\beta,k)$.
Let
\begin{equation*}
\begin{split}
A_2^{(2)}(\phi)&\equiv \lan \theta, \cK^{(2)}_\phi\theta\ran\\
A_3^{(2)}(\phi)&\equiv \lan \phi,\IE\lt[\cvtil(\cvtil(\theta))\rt]\ran.
\end{split}
\end{equation*}

Noting that
\begin{equation}
\label{cond1}
\begin{split}
&\mathbb{E}_t^\ep[V_s^\vas(x)\otimes\tilde{V}_s^\vas(y)] \\
&=\int\int [e^{ix\cdot
k}-1][e^{-iy\cdot k'}-1]
e^{-a{|k|}^{2\beta}|s-t|\vas^{-2}}
e^{-a{|k'|}^{2\beta}|s-t|\vas^{-2}} 
\hat{V}_t^\vas(dk)\otimes\hat{\tilde{V}}_t^{\vas*}(dk')\\
&\quad +
\int 
[e^{ix\cdot
k}-1][e^{-iy\cdot k}-1]
\left[1-e^{-2a{|k|}^{2\beta}|s-t|\vas^{-2}}\right] \cE_{K,L}(\alpha+\beta,k)
\,dk
\end{split}
\end{equation}
\commentout{
and
\begin{equation}
\label{cond2}
\begin{split}
&\mathbb{E}_t^\ep[\tilde{V}_s^\vas(x)\otimes\tilde{V}_s^\vas(y)] \\
&=\int e^{i(x-y)\cdot
k}\hat{\tilde{V}_t^\vas}(dk)\otimes
\hat{\tilde{V}_t^\vas}(dk)e^{-2a{|k|}^{2\beta}|s-t|\vas^{-2}}\\
&\quad +
\int e^{i(x-y)\cdot
k}
\left[1-e^{-2a{|k|}^{2\beta}|s-t|\vas^{-2}}\right] \cE_{K,L}(\alpha+2\beta,k)
\,dk.
\end{split}
\end{equation}
}
we then have
\be
\label{1.44}
f_2^\vas(t)=\frac{\vas^2}{2a^2}f''(t)\left[{\lan
\tvas,\cvtil(\theta)\ran}^2-A_2^{(2)}(\tvas)\right]
\ee
and similarly
\be
\label{1.45}
f_3^\vas(t)=\frac{\vas^2}{2a^2}f'(t)
\lt[\lan\tvas,\cvtil(\cvtil(\theta))\ran-A_3^{(2)}(\tvas)\rt].
\ee

\begin{prop}\label{prop:4}
$$ \lim_{\ep\to 0}\sup_{t<t_0} \mathbb{E}|f_2^\vas(t)|=0,\quad \lim_{\ep\to 0}\sup_{t<t_0}
\mathbb{E}|f_3^\vas(t)|=0. $$
\end{prop}
\begin{proof}
We have the bounds
\beqn
\sup_{t<t_0}\IE|f_2^\vas(t)|&\leq&
\sup_{t<t_0}\frac{\vas^2}{2a^2}\|f''\|_\infty
\|T_0\|_\infty^2\|\n\theta\|_\infty^2\lt[\int_{|x|<M}
\IE|\veptil|^2(x)\,dx+\int_{|x|<M}|\Gamma^{(2)}(x,x)|\,dx\rt]\\
&\leq&C_1\vas^2\\
\sup_{t<t_0}
\IE|f_3^\vas(t)|&\leq&
\sup_{t<t_0}
\frac{\vas^2}{2a^2}\|f'\|_\infty
\|T_0\|_\infty \lt[\|\n\theta\|_\infty\int_{|x|<M}\IE|\veptil|^2(x)\,dx
\rt.\\
&&\lt. \quad\quad\quad+
\|\theta\|_\infty \lt[\int_{|x|<M} \IE |\veptil|^2(x)\,dx\rt]^{1/2}
\lt[\int_{|x|<M} \IE |\n\veptil|^2(x)\,dx\rt]^{1/2}\rt]\\
&\leq& C_2\vas^2 K^{2-\alpha-2\beta};
\eeqn
both of them tend to zero.
\end{proof}
We have
\beqn
\cA^\vas f_2^\vas(t)&=&\frac{1}{a}
f''(t)\left[-
\lan\tvas, \cv(\theta)\ran\lan\tvas,\cvtil(\theta)\ran + A^{(1)}_2(\tvas)\right]
+ R_2^\vas(t)\\
\cA^\vas f_3^\vas(t)&=&\frac{1}{a}
f'(t)\left[-\lan
\tvas,\cv(\cvtil(\theta))\ran + A^{(1)}_3(\tvas)\right]+
R_3^\vas(t)
\eeqn
with
\beq
R_2^\vas(t)& =&\frac{f'''(t)}{2}\left[
\frac{\ep^2\tkappa}{2a^2}\lan \tvas, \Delta\theta\ran-\frac{\ep}{a^2}
\lan \tvas, \cv(\theta)\ran\right]
\left[\lan \tvas,\cvtil(\theta)\ran^2-A^{(2)}_2(\tvas)\right]\nonumber\\
&&\quad +
f''(t) \lan\tvas,\cvtil(\theta)\ran\left[
\frac{\tkappa\ep^2}{2a^2}\lan \tvas,\Delta\cvtil(\theta)\ran-
\frac{\ep}{a^2}\lan \tvas,\cv(\cvtil(\theta))\ran\right]\nonumber\\
&&\quad -
f''(t)\left[\frac{\tkappa\ep^2}{4a^2} \lan \tvas,\Delta G^{(2)}_{\theta}\tvas\ran
-\frac{\ep}{a^2}\lan \tvas,\cv(G^{(2)}_{\theta}\tvas)\ran\right]
\label{36}
\eeq
where 
$G_\theta^{(2)}$ denotes the operator
\[
G_\theta^{(2)}\phi\equiv \int G^{(2)}_{\theta,\theta}(x,y)\phi(y)\,dy,
\]
and similarly
\beqn
R^\ep_3(t)&=&f''(t)\left[\frac{\tkappa\ep^2}{4a^2}\lan\tvas,\Delta\theta\ran
-\frac{\ep}{2a^2}\lan\tvas,\cv(\theta)\ran\right]
\left[\lan\tvas,\cvtil(\cvtil(\theta))\ran-A^{(2)}_3(\tvas)\right]\\
&&\quad+f'(t)\left[\frac{\tkappa\ep^2}{4a^2}
\lan\tvas,\Delta\cvtil(\cvtil(\theta))\ran-
\frac{\ep}{2a^2}\lan\tvas,\cv(\cvtil(\cvtil(\theta)))\ran\right]\\
&&\quad-f'(t)\left[\frac{\tkappa\ep^2}{4a^2}\lan\tvas,\Delta
\IE[\cvtil(\cvtil(\theta))]\ran+\frac{\ep}{2a^2}\lan\tvas,\cv(\IE[\cvtil(\cvtil(\theta))])\ran\right].
\eeqn

\begin{prop}\label{prop:5}
\[
\lim_{\ep\to 0}\sup_{t<t_0} \mathbb{E} |R_2^\vas(t)|=0,\quad \lim_{\ep \to 0}
\sup_{t<t_0} \mathbb{E}
|R_3^\vas(t)|=0.
\]
\end{prop}
\begin{proof}
The argument is entirely analogous to that for Proposition~\ref{prop:4}.
The most severe term without the prefactor $\tkappa$
occurs in the expression for $R^\ep_3(t)$ and can be
bounded as
\beq
\ep\IE\left|\lan \tvas,\cv(\cvtil(\cvtil(\theta)))\ran\right| &\leq&
\ep\|T_0\|_\infty
\IE|\cv(\cvtil(\cvtil(\theta)))|\nonumber\\
\nonumber
&\leq&
C_1\ep\|T_0\|_\infty \left(\int_{|x|<M}\IE |\vep|^2\,dx\right)^{1/2}\times
\\
&&\left(\int_{|x|<M}\left\{
\IE\lt[|\veptil|^4\rt]\IE\lt[|\n^2\veptil|^4\rt]\right\}^{1/2}\,dx+
\int_{|x|<M}\IE\lt[|\n\veptil|^4\rt]\,dx\rt)^{1/2}
\eeq
by assumption. The right side of the above
tends to zero if either
\[
\alpha+2\beta>3
\]
or
\be
\label{47}
\alpha+2\beta=3,\quad\lim_{\ep\to 0}\ep
\sqrt{\log{K}}=0
\ee
or
\be
\label{48}
\alpha+2\beta<3,\quad
\lim_{\ep\to 0}\ep K^{3-\alpha-2\beta}=0
\ee
is satisfied.
The term involving $\ep\lan\tvas,\cv(G_\theta^{(2)}\tvas)\ran$ 
can be
similarly estimated.

The most severe term involving the prefactor $\tkappa$ occurs in
$R^\ep_3$ and can be bounded as
\beq
\nonumber
\tkappa\ep^2\IE\lt|\lan\tvas,\Delta\cvtil(\cvtil(\theta))\ran\rt|
&\leq& C\tkappa\ep^2\|T_0\|_\infty
\lt(\int_{|x|<M}\IE\lt[|\n^3\veptil|^2\rt]\rt)^{1/2}\\
&\sim&\left\{\begin{array}{ll}
\tkappa\ep^2,&\hbox{for}\,\,\alpha+2\beta>4\\
\tkappa\ep^2\sqrt{\log{K}},&\hbox{for}\,\,\alpha+2\beta=4\\
\tkappa\ep^2 K^{4-\alpha-2\beta},&\hbox{for}\,\,\alpha+2\beta<4
\end{array}\right.
\eeq
the right side of which tends to zero if either
\[
\alpha+2\beta>4
\]
or
\[
\alpha+2\beta=4,\quad\lim_{\ep\to 0}
\tkappa\ep^2\sqrt{\log{K}}=0
\]
or
\be
\label{51}
3<\alpha+2\beta<4,\quad\lim_{\ep\to 0}
\tkappa\ep^2 K^{4-\alpha-2\beta}=0
\ee
or
\[
2<\alpha+2\beta<3,\quad \lim_{\ep\to 0}
\tkappa\ep^2K^{4-\alpha-2\beta}=
\lim_{\ep\to 0}
\ep K^{3-\alpha-2\beta}=0.
\]
Note that for $\alpha+2\beta\leq 2$
the condition (\ref{47}) or (\ref{48}) implies
that
\[
\lim_{\ep\to 0} \ep^2K^{4-\alpha-2\beta} =0.
\]
\end{proof}

Set
\[
R^\vas(t) = A_4^\vas(t) - R_2^\vas(t) - R_3^\vas(t).
\]
It follows from Propositions 3 and 5 that
\[
\lim_{\ep \to 0}\sup_{t<t_0}\IE|R^\ep(t)|=0.
\]
Recall that
\beqn
M_t^\vas(\theta)&=&f^\ep(t)-\int^t_0 \cA^\ep f^\ep(s)\,ds\\
&=& f(t)-f_1^\vas(t)-f_2^\vas(t)-f_3^\vas(t)
-
\int_0^t\frac{\tkappa}{2}f'(t)\lan\tvas,\Delta\theta\ran\,ds\\
&& - \int_0^t\frac{1}{a}\left[f''(s)
A_2^{(1)}(T_s^\vas)+f'(s) A_3^{(1)}(T_s^\vas)\right]\,ds -\int_0^t
R^\vas(s)\,ds
\eeqn
is a martingale. Now that (\ref{38.2})-(\ref{42.2}) are satisfied
we can identify the limiting martingale to be
\begin{equation}
M_t(\theta)=f(t)-\int_0^t
\bigg\{f'(s)\left[\frac{\kappa_0}{2}\lan
T_s,\Delta\theta\ran+\frac{1}{a}\bar{A}_3^{(1)}(T_s)\right]
 + \frac{1}{a}f''(s)\bar{A}_2^{(1)}(T_s)\bigg\}\,ds
 \label{37}
\end{equation}
where
\[
\bar{A}_2^{(1)}(\phi)=\lim_{K\to\infty}{A}_2^{(1)}(\phi),
\quad \bar{A}_3^{(1)}(\phi)=\lim_{K\to\infty}{A}_3^{(1)}(\phi)
\]
(cf. (\ref{40.3}), (\ref{41.2})). 

Since $\lan\tvas,\theta\ran$ is uniformly bounded
\[
\lt|\lan\tvas,\theta\ran\rt|\leq \|T_0\|_\infty{\|\theta\|}_1
\]
we have the convergence of the second moment
\[
\lim_{\ep\to 0}
\IE\left\{
{\lan\tvas,\theta\ran}^2\right\}=\mathbb{E}\left\{ {\lan
T_t,\theta\ran}^2\right\}.
\]
Use $f(r) =r$ and $r^2$ in (\ref{37})
$$ M_t^{(1)}(\theta)=\lan T_t,\theta\ran -
\int_0^t \left[\frac{\kappa_0}{2}\lan T_s,\Delta\theta\ran +
\frac{1}{a}\bar{A}_3^{(1)}(T_s)\right]\,ds $$ is a martingale with the
quadratic variation
$$
\left[M^{(1)}(\theta),M^{(1)}(\theta)\right]_t=
\frac{2}{a}\int_0^t\bar{A}_2^{(1)}(T_s)\,ds=
\frac{2}{a}\int_0^t
\lan\theta,\bar{\cK}^{(1)}_{T_s}\theta\ran\,ds $$ where
$\bar{\cK}^{(1)}_{T_t}$ is a positive-definite operator given formally as
\be
\label{15.2}
\bar{\cK}^{(1)}_{T_t}\theta=
\int\theta(y) \nabla T_t(x)\cdot\bar{\Gamma}^{(1)}(x,y)\n T_t(y)
\,dy
\ee
(cf. \ref{15.2'}).
Therefore, $$
M_t^{(1)}=\sqrt{\frac{2}{a}}\int_0^t \sqrt{\bar{\cK}^{(1)}_{T_s}}dW_s $$ where
$W_s$ is a cylindrical Wiener process (i.e.
$dW_t(x)$ is a space-time white noise field)
and $\sqrt{\bar{\cK}^{(1)}_{T_s}}$ is the square-root of
the positive-definite operator given  in (\ref{15.2}).
From (\ref{40.3}) and (\ref{41.2}) we see that the limiting process $T_t$
is the distributional solution
to the martingale problem (\ref{13.2}) of
the It\^{o}'s equation 
\beqn
dT_t&=&
\left(\frac{\kappa_0}{2}\Delta +
\frac{1}{a}\bar{\cB}\right)
T_t\,dt+\sqrt{2a^{-1}\bar{\cK}^{(1)}_{T_t}}\,dW_t\\
&=&\left(\frac{\kappa_0}{2}\Delta +
\frac{1}{a}\bar{\cB}\right)
T_t\,dt+\sqrt{2}a^{-1/2}\n T_t\cdot d\bar{W}^{(1)}_t 
\eeqn
where the operator $\bar{\cB}$ is given by (\ref{56})
 and
$\bar{W}_t^{(1)}$ is the Brownian vector field with the spatial
covariance $\bar{\Gamma}^{(1)}(x,y)$.

\section{Proof of Theorem~\ref{thm2}}

As we let $L\to\infty$ along with $\ep\to 0$ the proof of
the uniform integrability of $\cA^\ep [f(t)-f_1^\ep(t)]$
(the first part of Proposition~3) breaks down. In this case,
we work with the perturbed test function
\[
f^\ep(t)=f(t)-f_1^\ep(t)+f_2^\ep(t)+f_3^\ep(t).
\]

\begin{prop}\label{prop:2'}
\be
\lim_{\ep\to 0}\sup_{t<t_0} \mathbb{E} |f_j^\vas(t)|=0,\quad
\lim_{\ep\to 0}\sup_{t<t_0} |f_j^\vas(t)|= 0
\quad \hbox{in probability},\quad\forall j=1,2,3.
\ee
\end{prop}
\begin{proof}
The argument for the case of $f^\ep_1(t)$ is the same as Proposition~2.
For $f^\ep_2(t)$ and $f^\ep_3(t)$ 
we have the bounds
\beqn
\sup_{t<t_0}\IE|f_2^\vas(t)|&\leq&
\sup_{t<t_0}\frac{\vas^2}{2a^2}\|f''\|_\infty
\|T_0\|_\infty^2\|\n\theta\|_\infty^2\lt[\int_{|x|<M}
\IE|\veptil|^2(x)\,dx+\int_{|x|<M}|\Gamma^{(2)}(x,x)|\,dx\rt]\\
&\leq&C_1\vas^2L^{2(\alpha+2\beta)-4}\\
\sup_{t<t_0}
\IE|f_3^\vas(t)|&\leq&
\sup_{t<t_0}
\frac{\vas^2}{2a^2}\|f'\|_\infty
\|T_0\|_\infty \lt[\|\n\theta\|_\infty\int_{|x|<M}\IE|\veptil|^2(x)\,dx
\rt.\\
&&\lt. \quad\quad\quad+
\|\theta\|_\infty \lt[\int_{|x|<M} \IE |\veptil|^2(x)\,dx\rt]^{1/2}
\lt[\int_{|x|<M} \IE |\n\veptil|^2(x)\,dx\rt]^{1/2}\rt]\\
&\leq& C_2\vas^2 L^{2(\alpha+2\beta)-4}
\eeqn
both of which vanish under the assumptions of the theorem.
Here we have used the fact that
\[
\int_{|x|<M}\IE|\veptil|^2(x)\,dx =O(L^{2(\alpha+2\beta)-4}),\quad L\to\infty.
\]
As for estimating $\sup_{t<t_0} |f_j^\vas(t)|, j=2,3$, we
can use 
\[
M^d\int_{|x|<M}|\veptil|^2(x)\,\,dx \quad \hbox{in place of}
\quad\int_{|x|<M} \IE |\veptil|^2(x)\,dx
\]
in the above bounds and obtain by assumption (cf.
(\ref{1.4''}), Remark 2) the desired estimate
which have a similar order of magnitude
with an additional factor of $1/\ep$
and a random constant possessing finite moments.
\end{proof}

We have
\beq
\cA^\ep f^\ep(t)&=&
\frac{\tkappa}{2}f'(t)\lan \tvas,\Delta\theta \ran
-\frac{1}{a} f''(t)A_2^{(1)}(\tvas)-\frac{1}{a}f'(t)A_3^{(1)}(\tvas)
+R^\ep_1(t)+R^\ep_2(t)+R^\ep_3(t)
\label{2.67}
\eeq
with
\be
R_1^\ep(t)=
\frac{\tkappa\vas}{2a}\left[f''(t)\lan\tvas,\Delta\theta\ran\lan\tvas,
\cvtil(\theta)\ran
 -f'(t)\lan\tvas,
  \Delta\cvtil(\theta)\ran\right]
   \ee
and $R_2^\ep(t), R_3^\ep(t)$ as before.

\begin{prop}\label{prop:7'}
\[
\lim_{\ep\to 0}\sup_{t<t_0}\IE|R^\ep_j(t)|=0,\quad
j=1,2,3.
\]
\end{prop}
\begin{proof}
The proof is similar to that of Proposition~5 with the additional
consideration due to $L\to\infty$. These additional terms
can all be estimated by 
\[
C_1\ep\int_{|x|<M}\IE\lt[\lt|\tilde{V}^\ep_t(x)\tilde{V}^\ep_t(x)
\rt|\rt]\,dx
\leq C_2\ep L^{2(\alpha+2\beta-2)}
\]
which tends to zero under the assumptions of the theorem.
\end{proof}

For the tightness
it remains to show
\begin{prop}
$\{\cA^\ep f^\ep\}$ are uniformly integrable.
\end{prop}
\begin{proof}
We shall prove that each term in the expression (\ref{2.67}) is
uniformly integrable. 

The first three terms are clearly bounded under the assumption
of $\alpha+\beta<2$. The last three terms can be estimated
as in Proposition~\ref{prop:7'} by
\[
C_1\ep\sup_{t<t_0}\int_{|x|<M}\lt|\tilde{V}^\ep_t(x)\tilde{V}^\ep_t(x)
\rt|
\]
whose second moment behaves like
$\ep^2 L^{4(\alpha+2\beta-2)}$, by the 4th order scale-invariance
property, and tends to zero.
\end{proof}

Now we have all the estimates needed to identify the limit
as in the proof of Theorem~\ref{thm1}.


\bigskip

{\large \bf Acknowledgments}\,\,
I thank L. Biferale and K. Gawedzki for stimulating
discussions on the nature of Lagrangian turbulent velocity during
the ``Developed Turbulence'' program, June 2002, at
The Erwin Schr\"{o}dinger International Institute
for Mathematical Physics, Vienna. I appreciate  the financial support
and hospitality
of ESI.
The research is supported in part by The Centennial Fellowship
from American Mathematical Society and a grant from U.S. National
Science Foundation, DMS-9971322.

\end{document}